%% file: 0-Main.tex
\documentclass[pdflatex,sn-apa]{sn-jnl}

\usepackage{fix-cm}
\usepackage{graphicx, calc}%
\usepackage{graphicx}
\usepackage{array}%
\usepackage{multirow}%
\usepackage{amsmath,amssymb,amsfonts}%
\usepackage{amsthm}%
\usepackage{mathrsfs}%
\usepackage[title]{appendix}%
\usepackage{xcolor}%
\usepackage{textcomp}%
\usepackage{manyfoot}%
\usepackage{booktabs}%
\usepackage{algorithm}%
\usepackage{algorithmicx}%
\usepackage{algpseudocode}%
\usepackage{listings}%
\usepackage{wrapfig}

\usepackage[utf8]{inputenc}%
\usepackage[T2A]{fontenc}%
\usepackage{caption} 
\usepackage[misc,geometry]{ifsym}
\usepackage{subcaption}
\usepackage[font=small]{caption, subcaption}
\usepackage{flushend}
\usepackage{pdflscape}
\usepackage{adjustbox}
\usepackage[english]{babel}

\usepackage{colortbl}
\usepackage{cleveref}
\usepackage{url}

\usepackage{adjustbox}
\usepackage{tcolorbox}
\tcbuselibrary{listings, skins, breakable}
\usepackage{cuted}
\usepackage{lipsum}

\definecolor{rank1}{HTML}{32A852}  
\definecolor{rank2}{HTML}{3CB371}  
\definecolor{rank3}{HTML}{48C774}  
\definecolor{rank4}{HTML}{66FF66}  
\definecolor{rank5}{HTML}{99FF99}  
\definecolor{rank6}{HTML}{CCFFCC}  
\definecolor{rank7}{HTML}{E6FFE6}  
\definecolor{rank8}{HTML}{F9FFF9} 
\usepackage[normalem]{ulem}
\useunder{\uline}{\ul}{}

\theoremstyle{thmstylethree}%

\begin{document}

\title[Article Title]{\centering From Reddit to Generative AI: Evaluating Large Language Models for Anxiety Support Fine-tuned on Social Media Data}






\author[1]{\fnm{Ugur} \sur{Kursuncu}}\email{ugur@gsu.edu}
\author[1]{\fnm{Trilok} \sur{Padhi}}\email{tpadhi1@student.gsu.edu} \equalcont{These authors contributed equally to this work.}
\author[2]{\fnm{Gaurav} \sur{Sinha}}\email{grsinha@uga.edu} \equalcont{These authors contributed equally to this work.}
\author[1]{\fnm{Abdulkadir} \sur{Erol}}\email{aerol1@student.gsu.edu} \equalcont{These authors contributed equally to this work.}
\author[3]{\fnm{Jaya Krishna} \sur{Mandivarapu}}\email{jmandivarapu@microsoft.com} \equalcont{These authors contributed equally to this work.}
\author[4]{\fnm{Christopher R.} \sur{Larrison}}\email{larrison@illinois.edu}

\affil[1]{\orgdiv{Georgia State University}, \city{Atlanta}, \state{GA}, \country{USA}}
\affil[2]{\orgdiv{University of Georgia}, \city{Athens}, \state{GA}, \country{USA}}
\affil[3]{\orgdiv{Microsoft}, \city{Atlanta}, \state{GA}, \country{USA}}
\affil[4]{\orgdiv{University of Illinois at Urbana-Champaign},
\state{IL}, \country{USA}}



\abstract{
The growing demand for accessible mental health support, compounded by workforce shortages and logistical barriers, has led to increased interest in utilizing Large Language Models (LLMs) for scalable and real-time assistance. However, their use in sensitive domains such as anxiety support remains underexamined. This study presents a systematic evaluation of LLMs (GPT and Llama) for their potential utility in anxiety support by using real user-generated posts from the \texttt{r/Anxiety} subreddit for both prompting and fine-tuning. Our approach utilizes a mixed-method evaluation framework incorporating three main categories of criteria: (i) linguistic quality, (ii) safety and trustworthiness, and (iii) supportiveness. Results show that fine-tuning LLMs with naturalistic anxiety-related data enhanced linguistic quality but increased toxicity and bias, and diminished emotional responsiveness. While LLMs exhibited limited empathy, GPT was evaluated as more supportive overall. Our findings highlight the risks of fine-tuning LLMs on unprocessed social media content without mitigation strategies. 
}


\maketitle

\section{Introduction}
\label{sec:introduction}
\input{1-Introduction}

\section{Related Work}
\label{sec:related-work}

\input{2-RelatedWork}

\section{Methods}
\label{sec:methodology}

\input{3-Methods}

\section{Results}
\label{sec:results}
\input{4-Results_Discussion}


\section{Discussion and Conclusion}
\label{sec:conclusion}

\input{5-Conclusion}

\bibliography{sn-bibliography}

\clearpage

\input{7-Appendix}

\end{document}

%% file: 1-Introduction.tex
Large Language Models (LLMs) have emerged as transformative technologies capable of generating human-like content, and are increasingly being explored for mental health applications. The critical shortage of mental health services due to workforce limitations and logistical barriers, especially in underserved areas designated by the Health Resources \& Services Administration (HRSA)\footnote{https://data.hrsa.gov/tools/shortage-area}, highlights the urgent need for accessible and scalable solutions. Traditional services often fail to address the diverse needs of individuals experiencing anxiety, prompting many, especially younger populations, to seek alternative emotional and psychological support online. While digital platforms offer immediate access, unregulated online interactions, including those with generative AI, may disseminate misleading information or inappropriate advice, potentially exacerbating anxiety symptoms \citep{tobias2021anxiety}.

Despite the great potential of generative AI to supplement mental health services, its deployment poses potentially significant risks. Unlike clinical practitioners, LLMs are not inherently equipped to manage emotionally complex or vulnerable conversations, which are critical to therapeutic relationships that create positive clinical outcomes \citep{rogers1957necessary, wampold2015important}. Misleading, insensitive, or toxic responses can easily trigger or escalate distress, particularly for people experiencing symptoms of anxiety, which manifests as excessive worry that impairs social and occupational functioning in a significant manner \citep{edition1980diagnostic}. Notable incidents, reported in popular media have shown that AI-driven interactions, in the absence of inadequate safeguards, have led to harmful consequences, including exacerbation of symptoms, misleading information, and in extreme cases, self-harm or suicide\footnote{https://www.nytimes.com/2024/10/23/technology/characterai-lawsuit-teen-suicide.html}. Despite growing public interest and their commercial deployment of AI virtual assistants for wellness, there remains a significant gap in rigorous, domain-specific evaluations of LLMs designed for mental health, specifically Anxiety.

To address this gap, we conducted a systematic, multidimensional evaluation of LLMs, including GPT-3.5 and Llama 2, using authentic user-generated data from the \texttt{r/Anxiety} subreddit. This data was used for both prompting and fine-tuning, allowing us to examine their performance under conditions that closely resemble real-world interactions, rather than relying on synthetic or artificial scenarios. This study makes the following three contributions: (i) evaluating LLMs utilizing anxiety-related posts from Reddit for prompting and fine-tuning, enhancing ecological validity; (ii) developing and applying a domain-specific multi-dimensional evaluation framework designed for anxiety; and (iii) analyzing the impact of fine-tuning with naturalistic anxiety-specific data.

Our evaluation framework is grounded in psychotherapy principles \citep{rogers1957necessary, wampold2015important} and dimensions from the Scales to Assess the Therapeutic Relation \citep{mcguire2007new}, and spans three categories: (i) linguistic quality (e.g., semantic coherence, readability, BERTScore, BLEURT, ROUGE); (ii) safety and trustworthiness (e.g., toxicity, harm, demographic bias); and (iii) supportiveness (e.g., empathy, reflection, sympathy). We incorporate both automated metrics and expert clinician ratings from licensed mental health professionals. Statistical analyses are conducted using Welch’s ANOVA, Games-Howell post-hoc tests, and Hedges’ g for effect sizes. 

Our findings reveal that fine-tuning on social media data improves accessibility and coherence. However, it introduces a higher likelihood of toxic or inappropriate content and diminishes the capacity for empathy and sympathy substantially. While GPT 3.5 outperforms Llama 2 in linguistic quality and safety, Llama 2 appears to be more empathetic. Both LLMs seemed to struggle with linguistic clarity and relevance after fine-tuning, highlighting the need for architectural adaptations or enhanced fine-tuning protocols. These findings highlight the significance of integrating mitigation strategies for toxicity, harm, and bias, and data curation in domain-specific fine-tuning for mental health applications. These insights provide timely guidance for researchers, developers, and mental health practitioners seeking to integrate generative AI technologies into their care systems.

%% file: 2-RelatedWork.tex

\subsection{LLMs in Mental Health Applications}

LLMs have emerged as promising tools to improve accessibility and scalability in mental health care, particularly through preliminary assessment, screening, and conversational support. Recent reviews have recognized this potential, including triage and patient education, and emphasized key limitations such as emotional misalignment, algorithmic bias, and lack of transparency in clinical or therapeutic contexts \citep{Hua2024LargeLM, Guo2024LargeLM}.

Domain-specific adaptations of LLMs have been developed to improve contextual grounding in mental health settings. Prior work has demonstrated the feasibility of mapping Reddit content to DSM-5 categories \citep{Gaur2018LetMT}, while models including \emph{PsychBERT} have shown improvements in classification tasks through fine-tuning on mental health-related social media data \citep{Vajre2021PsychBERTAM}. Similarly, \emph{MentaLLaMA} employed appraisal theory to improve interpretability of user mental states \citep{Yang2023MentaLLaMAIM}, and knowledge-infused frameworks have enhanced diagnostic summarization in clinical settings \citep{Manas2020KnowledgeInfusedAS}. Beyond diagnosis, LLMs were evaluated for their conversational competencies. For instance, the \emph{CounselingBench} benchmark revealed that even advanced models including GPT-4 struggle with sustaining empathy and ethical sensitivity across sessions \citep{Nguyen2024DoLL}. Appraisal-based evaluations further indicate that LLMs often misrepresent emotional variables such as controllability and changeability \citep{Yongsatianchot2023InvestigatingLL}. Concerns over prompt sensitivity, safety, and the potential for harmful outputs have also been well-documented \citep{Grabb2023TheIO, Metzler2022DetectingPH}.

Beyond model development, implementation efforts, such as CareBot, and human-AI alignment studies demonstrate additional challenges in trust calibration and interpretability \citep{Crasto2021CareBotAM, Elyoseph2024AssessingPI}. Collectively, prior work addresses key risks and opportunities of LLM-based mental health support, including prediction, interpretability, emotional reasoning, and safety, yet tends to evaluate LLM capabilities in isolation.

\subsection{Evaluation of Large Language Models} 
Prior work has increasingly recognized that LLM evaluation must extend beyond traditional metrics to include broader societal and ethical considerations and user-centered dimensions. Several frameworks have emerged advocating for multidimensional assessments incorporating fairness, correctness, and harm mitigation \citep{chang2023survey, guo2023evaluation, wang2024testing, saha2025large, Erol2025PlayingDA}. Reviews by \citet{minaee2024survey} and \citet{llmeval2024guidelines} further propose structured and reproducible evaluation protocols that emphasize contextual relevance and robustness to adversarial prompts.

However, these evaluations often overlook critical affective and relational dimensions, such as empathy, supportiveness, and emotional alignment, that are essential in mental health contexts. While some studies assess LLMs for diagnostic accuracy or linguistic quality \citep{roy2025diagnostic}, few provide comprehensive frameworks integrating therapeutic relevance, safety, and trustworthiness. A recent systematic review by \citet{Guo2024LargeLM} highlights this gap, noting the absence of multidimensional evaluations aligned with psychotherapy principles.

In contrast to this prior work, this study addresses three key existing gaps by: (i) utilizing naturalistic Reddit posts to enhance ecological validity and ground evaluations in authentic user concerns; (ii) developing and applying a domain-specific evaluation framework grounded in psychotherapy literature; (iii) examining the impact of fine-tuning to observe how domain adaptation influences model behavior in emotionally sensitive contexts. Through combined quantitative and expert qualitative analysis, we provide a rigorous, context-aware evaluation of LLMs for anxiety support.

%% file: 3-Methods.tex
In this study, we leverage naturalistic social media posts from the \texttt{r/Anxiety} subreddit to prompt and fine-tune LLMs, and evaluate their generated responses using a domain-specific framework based on three categories: (i) linguistic quality and understandability, (ii) safety and trustworthiness, and (iii) supportiveness. Grounded in psychotherapy common factors and the STAR measure, this framework provides a robust assessment of LLM responses in mental health contexts \citep{guo2023evaluation, minaee2024survey}. We outline the quantitative evaluation approach first, followed by qualitative assessment.

\begin{figure}[h]
  \centering
  \includegraphics[width=13cm]{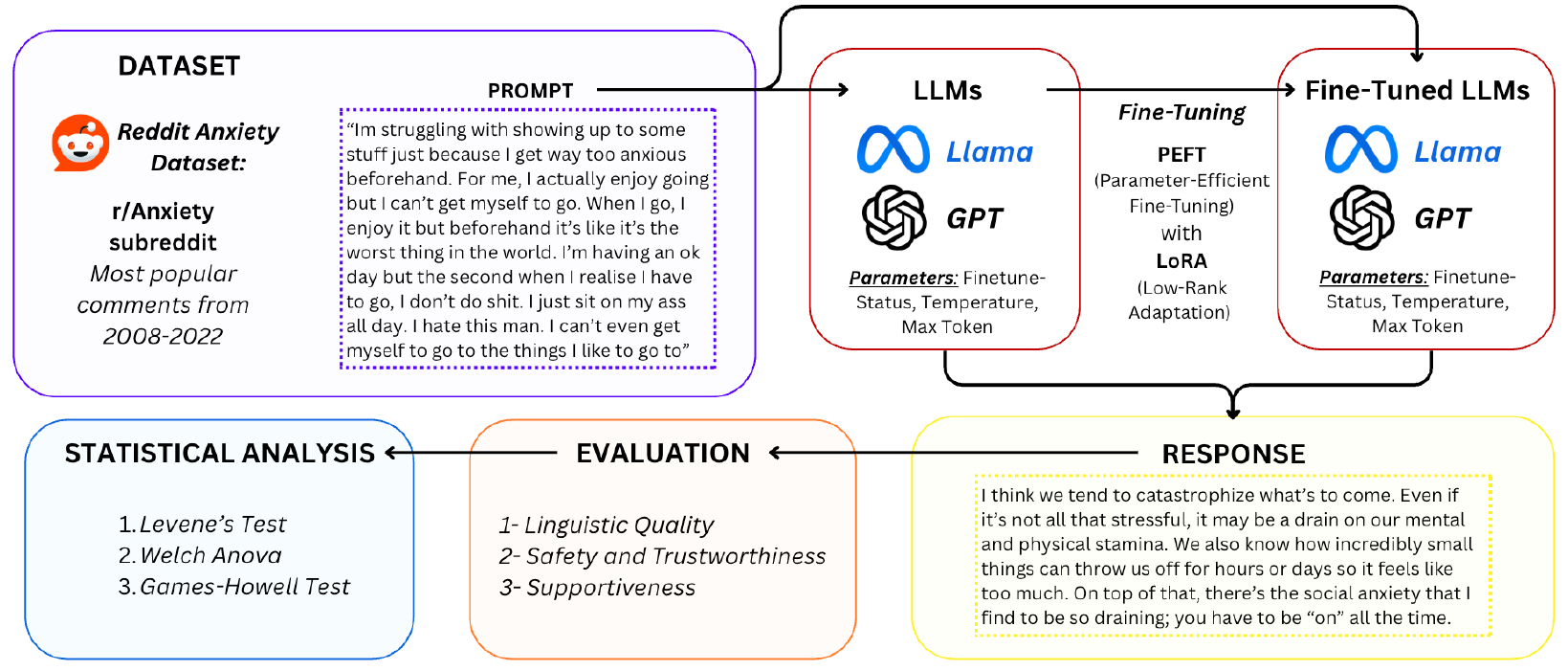}
  \caption{Our approach for assessing LLMs for anxiety support, utilizing real user-generated data from the \texttt{r/Anxiety} subreddit for both prompting and fine-tuning Llama 2 and GPT 3.5. Subsequently, the generated responses are systematically evaluated across three key categories: linguistic quality, safety and trustworthiness, and supportiveness, employing a combination of automated metrics and qualitative human evaluations by mental health professionals. }
  \label{fig:framework}
\end{figure}

\begin{figure}[h]
  \centering
  \includegraphics[width=13cm]{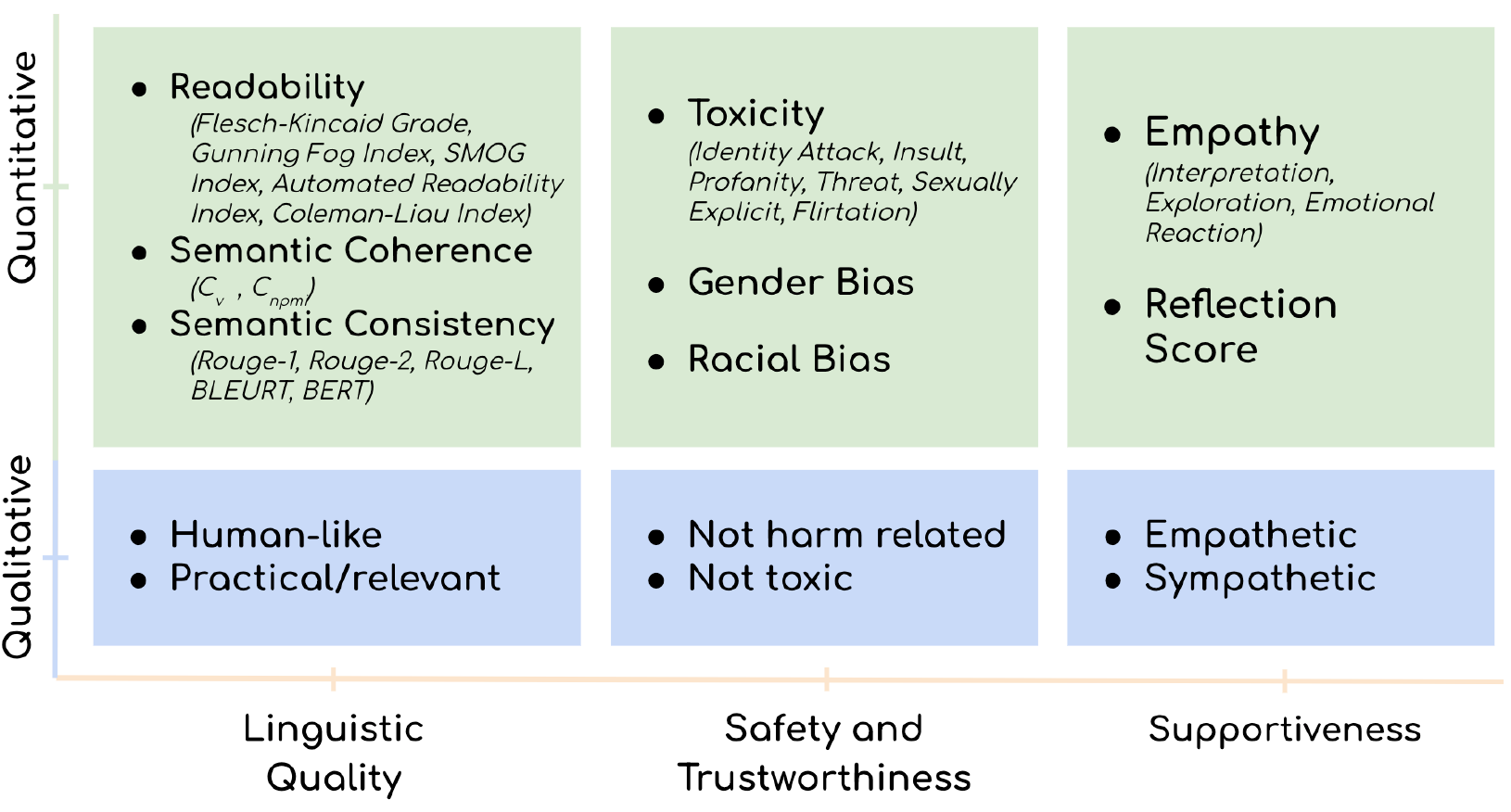}
  \caption{Evaluation framework overview. Metrics are grouped by three core dimensions—Linguistic Quality, Safety and Trustworthiness, and Supportiveness—and categorized as either quantitative or qualitative.}
  \label{fig:evaluation}
\end{figure}

\subsection{Dataset} 
For this study, we employed authentic user-generated data from the \texttt{r/Anxiety} subreddit, specifically utilizing main posts and comments authored by anonymous individuals discussing genuine anxiety-related concerns \citep{sharma2018mental}. We utilized such authentic posts to prompt LLMs to ensure the generated responses reflect realistic interactions rather than synthetic or artificial scenarios. For fine-tuning LLMs, we created a dataset, consisting of high-quality prompt-response pairs, aimed at providing supportive and insightful responses to the posts from individuals experiencing anxiety. The data initially comprised 544,377 posts and 1,535,328 comments. We removed deleted posts and the posts without titles, as titles provide essential context for understanding the content, yielding 396,272 posts. To focus on the quality interactions, we excluded self-comments and retained only posts with at least one comment scored greater than four by the contributors (upvotes – downvotes) \citep{reddit_selfpromotion, reddit_vote_cheating}, resulting in 26,450 posts. This final dataset was split into 21,000 for fine-tuning and 5,000 for testing. 

\subsection{LLM Selection, Fine-tuning and Training Objective}
We selected LLMs from two leading model families, Llama and GPT, based on their broad adoption, proven technical capabilities, extensive usage, and accessibility for various research and practical applications \citep{kapoor2024societal, touvron2023llama, brown2020language}. Specifically, we utilized Llama 2 (7B) and GPT 3.5 Turbo as representative models of open-source and proprietary systems, respectively, allowing comparative analysis in sensitive domains such as mental health. Their moderate parameter sizes balance performance with computational efficiency, making them suitable for deployment in resource-constrained settings \citep{tan2024democratizing}. We acknowledge that due to the rapid pace of AI research and development, newer versions of these models may become available by the time of publication. However, the findings and framework presented here remain highly relevant, offering a replicable approach for evaluating future models in mental health contexts. 

To enable domain adaptation, both LLMs were fine-tuned using a dataset of prompt-response pairs curated from the \texttt{r/Anxiety} subreddit. Fine-tuning was designed to optimize the models' ability to generate linguistically coherent, safe, and emotionally supportive responses to anxiety-related user input. The fine-tuning process minimizes the negative log-likelihood (NLL) loss over $N$ training examples:
\[
\mathcal{L}(\theta) = -\frac{1}{N} \sum_{i=1}^{N} \log P(C_i \mid P_i)
\]
where $P_i$ is the user-generated prompt, $C_i$ is the corresponding human response, and $\theta$ represents the model parameters. $P(C_i \mid P_i)$ denotes the probability of generating $C_i$ given $P_i$. Llama 2 (7B) was fine-tuned using Parameter-Efficient Fine-Tuning (PEFT) with LoRA (Low-Rank Adaptation) to reduce the number of trainable parameters while maintaining model performance. LoRA injects trainable low-rank matrices into the attention layers of the transformer architecture, thereby reducing memory and compute requirements during fine-tuning \citep{hu2022lora}.

GPT-3.5 Turbo was fine-tuned via OpenAI’s fine-tuning API using prompt-completion pairs, following its constrained tuning schema and batch training setup. We experimented with different parameters for inference (e.g., temperature, top-$k$), but observed no significant effect on the evaluation metrics; hence, default values were used.

\subsection{Quantitative Evaluation Process}
A domain-specific evaluation framework allows us to identify patterns beyond standard predictive accuracy measures. A multi-dimensional framework highlights the importance of designing an evaluation approach that reflects the actual expectations of users in sensitive domains. In mental health, a supportive response should not only be accurate, but also be emotionally attuned and non-judgmental \citep{rogers1957necessary, wampold2015important}. In addition to linguistic quality and safety, the supportiveness measures, such as empathy, sympathy, and reflection, provide interpretive richness to the analysis and align model performance with real-world needs.

\subsubsection{Linguistic Quality}	Linguistic quality was assessed to evaluate the accuracy, contextual relevance, and clarity of LLM-generated responses to authentic anxiety-related prompts. To evaluate the linguistic quality of model outputs, we employ readability, semantic coherence, and semantic consistency metrics. Readability metrics, including Flesch-Kincaid Grade Level (FKG), Gunning Fog Index (GFI), SMOG Index (SI), Automated Readability Index (ARI), and Coleman-Liau Index (CLI) assess how easily a general audience can understand the text. Semantic coherence metrics, such as $C_v$, $C_{npmi}$, assess the coherence of topics and the overall thematic structure within a response by analyzing word co-occurrence patterns. Semantic consistency such as BERT Score \citep{zhang2019bertscore}, BLEURT \citep{sellam2020bleurt}, and ROUGE \citep{lin2004rouge} evaluate how well the generated content aligns with reference responses in both structure and meaning. These metrics provide complementary perspectives on fluency, clarity, and semantic coherence. Table \ref{tab:metric-explanations} summarizes these metrics, their evaluation methods and their brief description, and Appendix \ref{sec:linguistic-quality} provides detailed definitions and computational formulations of these metrics. We employed three established metrics: ROUGE, BLEURT, and BERTScore. ROUGE measures lexical overlap between model outputs and reference responses curated by domain experts, emphasizing recall of anxiety-relevant terminology. BLEURT, a fine-tuned transformer-based model, evaluates semantic and syntactic alignment, capturing fluency and naturalness. BERTScore uses contextual embeddings from BERT to compute semantic similarity, providing a deeper measure of how meaningfully the model captures user intent and domain-specific nuance. 

\subsubsection{Safety \& Trustworthiness.}	This dimension evaluates harmful or inappropriate elements in generated responses that may undermine trust or emotional safety in mental health contexts \citep{mccullough2020trustworthiness}. We focus on three key risks: toxicity, demographic bias, and unsolicited medical advice . We use the Perspective API \citep{perspective_api} to compute probabilistic scores across six categories (Toxicity, Severe Toxicity, Insult, Profanity, Identity Attack, Threat) to assess the likelihood of harmful or offensive language. To further evaluate demographic bias, we utilized GenBit Score \citep{sengupta2021genbit}, and in-context impersonation \citep{salewski2023context}. GenBit Score \citep{sengupta2021genbit}, which measures diversity in language use, with higher values indicating less repetition and more varied language use, an important factor in detecting repetitive or biased patterns. Demographic bias is assessed via gender and racial cues. Gender bias is quantified using the Average Bias Conditional Absolute Score (ABCAS) derived from GenBit. For racial bias, we implement in-context impersonation, introducing identity-specific variations in prompts to uncover disparities in model behavior. To detect unsolicited advice, we solely rely on qualitative human evaluation due to lack of reliable quantitative method. This is especially important in peer-like interactions, where the goal is emotional validation rather than clinical diagnosis. Table \ref{tab:metric-explanations} summarizes their purposes under the Bias and Toxicity categories, and full definitions and computational details are provided in Appendix \ref{sec:safety-trustworthiness}. Together, these metrics offer a comprehensive view of safety and trustworthiness in LLM responses tailored for anxiety support.

\subsubsection{Supportiveness.} To evaluate supportiveness in LLMs, we adopt the theoretically grounded empathy framework proposed by \citet{sharma2020computational}, which decomposes empathy into three distinct mechanisms: \textit{emotional reaction}, \textit{interpretation}, and \textit{exploration}. These dimensions capture different ways in which support agents can express empathy in text-based mental health communication. Emotional reaction refers to affective acknowledgments of the user’s emotional state (e.g., “That must be really difficult”), interpretation involves making sense of the user’s experience or intentions (e.g., “It seems like you’re worried about…”), and exploration reflects the agent’s attempt to elicit more information or deepen engagement (e.g., “Can you tell me more about that?”). A summary of metric definitions is provided in Table \ref{tab:metric-explanations}, with further details in Appendix \ref{sec:supportiveness}.

To incorporate this empathy framework into our model, we modified the training dataset to reflect these empathic behaviors. Each user message was paired with one of the three empathy categories: \texttt{EMOTIONAL\_REACTION}, \texttt{INTERPRETATION}, or \texttt{EXPLORATION}. Please refer Appendix~\ref{sec:empathy} for the prompt and how we evaluated empathy scores using this prompt.

The data was restructured into a multi-task format, where each instance consisted of a user input, a task label, and a corresponding model-generated response. We curated this dataset using publicly available empathy classification corpora and supplemented it with additional in-house annotations to ensure coverage across all categories. The model was trained using a multi-task fine-tuning setup on \textbf{Llama 3.1}, with each task preceded by a task-specific prefix (e.g., \texttt{<INTERPRETATION>}). A shared encoder-decoder architecture was used, and training was guided by a weighted cross-entropy loss to address class imbalance, particularly in lower-resource categories like \texttt{REFLECTION} and \texttt{EXPLORATION}.

For the \texttt{REFLECTION} metric, we adopted the scoring methodology introduced in the PAIR model by \citet{Min2022PAIRPM}. PAIR uses a prompt-aware margin ranking approach to evaluate counselor reflections in motivational interviewing settings. Specifically, it measures how well a response mirrors the meaning of the user’s input while demonstrating meaningful lexical and syntactic transformation. The model is trained to distinguish high-quality reflective responses from shallow restatements or irrelevant output. We applied a version of this scoring approach to assess the paraphrastic quality of model responses in our evaluation set. Together, the empathy and reflection scores provide a multidimensional assessment of supportiveness—capturing both the emotional and cognitive alignment of the model's behavior. This methodology allows us to go beyond binary correctness judgments and evaluate whether models are capable of simulating nuanced, human-centered behaviors like emotional resonance, paraphrastic understanding, and conversational engagement—skills that are vital for building safe and effective mental health support systems.

\subsection{Qualitative Human Evaluation Process}
A total of 1,400 model responses, generated across different configuration settings, were randomly selected and evaluated by a team led by a professional social worker (S) with 16 years of practice experience. The team employed  the same evaluation matrix used for quantitative evaluation: linguistic quality (such as readability and authentic human-like response), safety and trustworthiness (harm-related contents or use of toxic, offensive content and swear words), and supportiveness (empathy and sympathy). First, a codebook (available upon request) was developedwith brief descriptions of each of the three themes along with  examples. This codebook was vetted by the team leader of quantitative evaluation. Next, two evaluators with practical experience working with clients in clinical mental health settings, were trained to  usee the codebook to assess 1,400 LLMs responses randomly selected (350 each for the two tuned and two untuned models). After the initial training, practice sessions were held to address to ensure consistent understanding of the metrics. Finally, the two evaluators independently evaluated the LLMs responses. To resolve disagreements, a third, evaluator, with the same qualifications, was employed to independently evaluate the same set of responses. In cases of disagreement, the majority decision (i.e., two out of three evaluators agreeing) was used to determine the final classification for each response.

\begin{table*}[hbt!]\centering
\scriptsize
\resizebox{13cm}{!}{
    \begin{tabular}{p{0.4cm}p{1.4cm}p{3.2cm}p{8.5cm}}\toprule
    & \textbf{Metric} & \textbf{Evaluation Method} & \textbf{Brief Description} \\\midrule
    \multirow{18}{*}{\textbf{LQ}} & \multirow{5}{*}{Readability} 
    & Flesch-Kincaid Grade & Measures reading level based on sentence length and syllables per word. \\
    & & Gunning Fog Index & Estimates the years of education needed to understand the text. \\
    & & SMOG Index & Assesses readability based on the number of complex words. \\
    & & Automated Readability Index (ARI) & Calculates readability using characters per word and words per sentence. \\
    & & Coleman-Liau Index & Uses characters instead of syllables to assess readability level. \\
    \cmidrule(l){2-4}
    & \multirow{4}{*}{\shortstack[l]{Semantic \\ Coherence}} 
    & $C_v$ & Measures topic coherence using a sliding window and cosine similarity. \\
    & & $C_{npmi}$ & Normalizes pointwise mutual information for better coherence comparison. \\

    \cmidrule(l){2-4}
    & \multirow{6}{*}{\shortstack[l]{Semantic \\ Consistency}} & BLEURT & Uses pretrained models to evaluate fluency and semantic similarity. \\
    & & ROUGE & Focuses on recall-based n-gram overlap, commonly used in summarization. \\
    & & BERTScore & Compares contextual embeddings from BERT to measure similarity. \\
    
    \midrule
    \multirow{10}{*}{\textbf{ST}} & \multirow{4}{*} {Bias} & GenBit Score \citep{sengupta2021genbit} & Measures and mitigates gender bias \\
    & & Racial Bias \citep{salewski2023context} & In-context impersonation, introducing identity-specific variations in prompts to uncover disparities in model behavior\\
    
    \cmidrule(l){2-4}
    & \multirow{6}{*}{Toxicity} & Perspective API \citep{perspective_api} & \texttt{Toxicity:} Rude or disrespectful comments. \newline
    \texttt{Insult:} Offensive or insulting comments. \newline
    \texttt{Profanity:} Swear words or obscene language. \newline
    \texttt{Identity Attack:} Hateful remarks targeting identity. \newline
    \texttt{Severe Toxicity:} Extremely hateful or aggressive content. \newline
    \texttt{Threat:} Intent to harm or threaten. \\
    
    \midrule
    \multirow{6}{*}{\textbf{Sp}} & \multirow{2}{*}{Empathy} & EPITOME \citep{sharma2020computational} & Empathy is assessed through \texttt{interpretation}, {emotional reaction}, and {exploration} of the seeker’s emotions and experiences. \\
    \cmidrule(l){2-4}
    & \multirow{2}{*}{Reflection} & PAIR \citep{perez2022pair} & A reflection score evaluates how effectively a counselor’s response demonstrates understanding
    and affirmation of a client’s experiences and concerns. \\
    
    \bottomrule
    \end{tabular}
}
\caption{This table outlines the key metrics, their evaluation method, and brief description used to evaluate metrics. For detailed definitions, scoring methods, and rationale behind these metrics, refer to Appendix \ref{sec:linguistic-quality}, and \ref{sec:safety-trustworthiness}, \ref{sec:supportiveness}. Abbreviations: LQ: Linguistic Quality, ST: Safety and Trustworthiness, Sp: Supportiveness.}
\label{tab:metric-explanations}
\end{table*}

\subsection{Statistical Analysis}
To evaluate differences in model performance across metrics in each category, we first applied Levene’s test to assess homogeneity of variance, which indicated unequal variances among subgroups. As a result, we employed Welch’s ANOVA \citep{welch1951comparison} for group comparisons, as it is robust to heteroscedasticity and unequal sample sizes. For post-hoc pairwise comparisons, we used the Games-Howell test \citep{games1976pairwise}, which accommodates both unequal variances and sample sizes. To quantify practical significance, we calculated Hedges’ g as a measure of effect size, offering improved accuracy over Cohen’s d in cases with small or unequal groups. We then assigned ranking scores to model configurations based on statistically significant differences and corresponding effect sizes. This dual approach allowed us to prioritize models that achieved statistical significance and demonstrated meaningful performance gains. This layered analysis offers a robust comparison framework across linguistic quality, safety and supportiveness, supportiveness, accounting for both statistical and practical relevance.

%% file: 4-Results_Discussion.tex
The results of our analysis are presented in Table~\ref{tab:results}, the effects of fine-tuning vary considerably between models and evaluation metrics. The following subsections provide a detailed breakdown of these results for each category. For detailed results, refer to Appendix~\ref{sec:detailed-results}.

\begin{table*}[hbt!]\centering
\scriptsize
\resizebox{13cm}{!}{
    \begin{tabular}{p{0.4cm}p{1.7cm}p{4.5cm}p{1.8cm}p{1.7cm}p{1.5cm}p{1.5cm}}\toprule
    & Metric & Evaluation Method & GPT-3.5-FT & Llama 2-FT & GPT-3.5 & Llama 2 \\
    \midrule
    \multirow{14}{*}{\textbf{LQ}} & \multirow{5}{*}{Readability} & Flesch-Kincaid Grade* & \cellcolor[HTML]{d9ead3}\textbf{1(-3)} & 3(1) & 2(-1) & 4(3) \\
    & & Gunning Fog Index* & \cellcolor[HTML]{d9ead3}\textbf{1(-3)} & 2(1) & 2(1) & 2(1) \\
    & & SMOG Index* & \cellcolor[HTML]{d9ead3}\textbf{1(-3)} & 3(1) & 2(-1) & 4(3) \\
    & & Automated Readability Index* & \cellcolor[HTML]{d9ead3}\textbf{1(-3)} & 3(2) & 2(-1) & 3(2) \\
    & & Coleman-Liau Index* & \cellcolor[HTML]{d9ead3}\textbf{1(-3)} & 2(0) & 2(0) & 4(3) \\
\cmidrule(l){2-7}
& \multirow{2}{*}{\shortstack[l]{Semantic \\ Coherence}} & $C_v$ & \cellcolor[HTML]{d9ead3}\textbf{1(2)} & 3(1) & \cellcolor[HTML]{d9ead3}\textbf{1(2)} & 2(0) \\
    & & $C_{npmi}$ & 4(0) & \cellcolor[HTML]{d9ead3}\textbf{1(3)} & 3(1) & 2(2) \\
\cmidrule(l){2-7}
& \multirow{7}{*}{\shortstack[l]{Semantic \\ Consistency}} & Rouge-1 & \cellcolor[HTML]{d9ead3}\textbf{1(1)} & 3(0) & \cellcolor[HTML]{d9ead3}\textbf{1(1)} & 3(0) \\
    & & Rouge-2 & \cellcolor[HTML]{f4cccc}0 & \cellcolor[HTML]{f4cccc}0 & \cellcolor[HTML]{f4cccc}0 & \cellcolor[HTML]{f4cccc}0 \\
    & & Rouge-L & \cellcolor[HTML]{f4cccc}0 & \cellcolor[HTML]{f4cccc}0 & \cellcolor[HTML]{f4cccc}0 & \cellcolor[HTML]{f4cccc}0 \\
    & & BLEURT & 2(1) & 4(0) & \cellcolor[HTML]{d9ead3}\textbf{1(3)} & 2(1) \\
    & & BERT Precision & 2(2) & 4(0) & \cellcolor[HTML]{d9ead3}\textbf{1(3)} & 3(1) \\
    & & BERT Recall & \cellcolor[HTML]{f4cccc}0 & \cellcolor[HTML]{f4cccc}0 & \cellcolor[HTML]{f4cccc}0 & \cellcolor[HTML]{f4cccc}0 \\
    & & BERT F1 & \cellcolor[HTML]{d9ead3}\textbf{1(2)} & 3(0) & \cellcolor[HTML]{d9ead3}\textbf{1(2)} & 3(0) \\
\midrule
\multirow{9}{*}{\textbf{ST}} & \multirow{2}{*}{Bias} & GenBit Scores* & 2(1) & 3(2) & \cellcolor[HTML]{d9ead3}\textbf{1(0)} & 3(2) \\
& & Racial Bias* & - & - & - & - \\
\cmidrule(l){2-7}
& \multirow{6}{*}{Toxicity} & Toxicity* & 3(2) & 4(3) & \cellcolor[HTML]{d9ead3}\textbf{1(0)} & 2(1) \\
    & & Insult* & 3(2) & 4(3) & \cellcolor[HTML]{d9ead3}\textbf{1(0)} & 2(1) \\
    & & Profanity* & 3(2) & 4(3) & \cellcolor[HTML]{d9ead3}\textbf{1(0)} & 2(1) \\
    & & Identity Attack* & 3(2) & 4(3) & \cellcolor[HTML]{d9ead3}\textbf{1(0)} & \cellcolor[HTML]{d9ead3}\textbf{1(0)} \\
    & & Severe Toxicity* & 3(2) & 3(2) & \cellcolor[HTML]{d9ead3}\textbf{1(0)} & 2(1) \\
    & & Threat* & 3(2) & 4(3) & \cellcolor[HTML]{d9ead3}\textbf{1(0)} & 2(1) \\
\midrule
\multirow{4}{*}{\textbf{Sp}} & Reflection
& Reflection Scores & 4(0) & 3(1) & \cellcolor[HTML]{d9ead3} \textbf{1(3)} & 2(2) \\
\cmidrule(l){2-7}
& \multirow{3}{*}{Empathy}
& Interpretation &3(1) &4(0) &2(2) & \cellcolor[HTML]{d9ead3} \textbf{1(3)} \\  
& &Emotional Reaction & \cellcolor[HTML]{d9ead3} \textbf{1(1)} & \cellcolor[HTML]{d9ead3} \textbf{1(1)} & \cellcolor[HTML]{d9ead3} \textbf{1(1)} & 4(0) \\  
& &Exploration &2(2) &3(1) &\cellcolor[HTML]{d9ead3} \textbf{1(3)} &4(0) \\  
\bottomrule
\end{tabular}
}
\caption{Ranking of model performance across various evaluation metrics. Each cell presents the \textit{“rank (ranking score)”} for a given model, where Rank 1 indicates the best performance. Metrics marked with an asterisk (*) denote that lower scores are better, and ranks are assigned accordingly. All other metrics are ranked with higher scores being better. Rankings are derived based on Welch’s ANOVA and Games-Howell post-hoc tests, ensuring statistically significant comparisons among the four model variants: GPT-3.5 Finetuned, GPT-3.5, Llama 2 Finetuned, and Llama 2. Equal scores share the same rank. Green shading indicates the best-performing model(s), while red indicates no statistically significant difference from other models. For detailed results table, refer to Appendix~\ref{tab:detailed-results}. Abbreviations: LQ: Linguistic Quality, ST: Safety and Trustworthiness, Sp: Supportiveness.}\label{tab:results}
\end{table*}

\subsection{Linguistic Quality}
A core aim of this study is to assess how LLMs perform in generating responses to anxiety-related authentic prompts from the Anxiety subreddit rather than synthetic inputs. We examine linguistic quality across three subcategories: Readability, Semantic Coherence, and Semantic Consistency (Textual Alignment). 

\subsubsection{Readability} Fine-tuning LLMs with anxiety-related user-generated data significantly improved their readability across nearly all metrics, including the Flesch-Kincaid Grade, Gunning Fog Index, SMOG Index, Automated Readability Index (ARI), and Coleman-Liau Index. GPT 3.5 consistently generated the most accessible responses, achieving the lowest complexity scores post-fine-tuning. For instance, GPT's readability and accessibility significantly improved after fine-tuning, with the Flesch-Kincaid Grade decreasing from 8.72 to 6.05 (30.62\% improvement), Gunning Fog Index from 11.21 to 8.47 (24.44\%), SMOG Index from 8.17 to 5.83 (28.64\%), ARI from 10.2 to 6.73 (34.02\%), and the Coleman-Liau Index from 7.55 to 5.43 (28.08\%). These consistent reductions across metrics suggest enhanced accessibility, simpler sentence structure, and reduced linguistic complexity—critical for engaging users experiencing anxiety. On the other hand, Llama 2 showed modest improvements after fine-tuning, with the Flesch-Kincaid Grade decreasing from 10.28 to 9.64 (6.23\% improvement), the Gunning Fog Index from 11.58 to 11.19 (3.37\%), the SMOG Index from 12.88 to 10.54 (18.17\%), and the Coleman-Liau Index from 11.68 to 9.16 (21.58\%). However, the ARI increased from 12.4 to 13.34 (7.58\% worse), suggesting that while some metrics reflect enhanced accessibility, the overall complexity of the generated responses may have increased slightly. 

\subsubsection{Semantic Coherence} Both LLMs showed improvements after fine-tuning, with GPT slightly outperforming Llama 2 in $C_v$. Specifically, GPT achieved a $C_v$ score of 1.0 in both fine-tuned and non-fine-tuned settings, while Llama 2 improved from 0.80 to 0.91 (13.75\%) after fine-tuning. Despite this improvement, Fine-tuned and non-fine-tuned GPT’s coherence remained 9.89\% and 25\% higher than Llama 2’s respective versions. Based on NPMI, fine-tuned versions of both LLMs produced higher coherence scores, where GPT’s CNPMI dropped from 0.16 to -0.22 (-237.5\%) and Llama 2’s score decreased from 0.13 to -0.10 (-176.92\%), suggesting that while fine-tuning improves some coherence dimensions, it may compromise global topical alignment in complex conversations. 

\subsubsection{Semantic Consistency (Textual Alignment)} Fine-tuned models LLMs demonstrated stronger consistency in preserving semantic alignment with reference responses. The results show modest but consistent improvements across ROUGE-1 and ROUGE-L, and particularly in BLEURT, where GPT improved by 27.2\% after fine-tuning. Llama 2 showed marginally higher ROUGE-2, suggesting better bigram-level contextual alignment in isolated cases. BERTScore results showed that GPT maintained better precision (i.e., more accurate token-level matches), although recall remained low across all models. Llama 2 provided poor performance on both precision and recall, suggesting that its generated content, even when fine-tuned, was more prone to omissions and semantic drift. 

\paragraph{\textit{Takeaway:}} Fine-tuning with authentic anxiety-related data improves readability, accessibility and fluency, particularly for GPT 3.5, making responses more suitable for users with various literacy levels and emotional needs. Fine-tuning yields trade-offs in semantic coherence; while it provides slight coherence gains, it diminishes global topic alignment (NPMI) and recall, highlighting the need for careful calibration of fine-tuning strategies when applying LLMs in real-world mental health contexts. Fine-tuning improved overall semantic alignment for both LLMs, with GPT indicating stronger ability to produce contextually relevant and textually consistent responses to anxiety-related prompts.

\subsection{Safety and Trustworthiness}
Fine-tuning LLMs with user-generated anxiety-related data revealed important trade-offs in safety and trustworthiness. 

\subsubsection{Bias} GPT models exhibited substantially lower bias, while fine-tuning increased GPT’s gender bias score from 0.17 to 0.37 (117.65\%), still remained less biased than both fine-tuned and non-fine-tuned Llama 2 (0.51 in both cases). These findings point to GPT’s more robust baseline behavior and highlight the need for additional approaches to adapt Llama 2, such as adversarial training. The relative difference in gender bias between Llama 2 to GPT was 27.45\% in the fine-tuned and 66.67\% in the non-fine-tuned settings, respectively. 

\subsubsection{Toxicity} Fine-tuning increased toxicity scores in both LLMs, but the effect was much more substantial for GPT, especially for insult (+200\%) and profanity (+350\%), while toxicity scores of GPT still remained lower than Llama 2. Llama 2 also showed increased toxicity after fine-tuning, but the increase was smaller at 90\%. In particular, all toxicity metrics remained below 0.5, suggesting that fewer than half of users are likely to perceive the content as harmful. While non-fine-tuned GPT was less toxic than fine-tuned Llama 2 across most metrics, fine-tuned Llama 2 consistently performed better, suggesting it appears to have learned less harmful linguistic traits from the fine-tuning data, although toxicity still increased.

\paragraph{\textit{Takeaway:}} Fine-tuning LLMs on authentic anxiety-related data can enhance contextual relevance and emotional alignment but also substantially increases the risk of generating toxic content, especially in GPT. Llama 2 was comparatively less affected, suggesting potential benefits of its open-source architecture for fine-tuning. Raw user-generated social media data can be often unfiltered, which may risk for safety unless it is preprocessed for safety before training.

\subsection{Supportiveness}
Supportiveness was evaluated using a combination of empathy-related dimensions, including Interpretation, Emotional Reaction, and Exploration, alongside Reflection Scores, which collectively reflect a model’s capacity for emotional understanding, engagement, and human-centered dialogue.

\subsubsection{Empathy}
Among all models, the non-fine-tuned Llama 2 (NFT-Llama 2) demonstrated the highest score on the \emph{Interpretation} metric (1.67), indicating a strong tendency to infer and restate user perspectives. However, this may not necessarily reflect deeper empathy, as high interpretation scores in NFT models could stem from overconfident or verbose outputs rather than genuine understanding. In contrast, fine-tuned GPT (FT-GPT) scored the lowest on this dimension (1.37), suggesting that fine-tuning may suppress elaborative or interpretive behaviors in favor of more direct and possibly conservative responses. Both Llama 2 and GPT models showed a decrease in Interpretation following fine-tuning, with GPT exhibiting a smaller decline (-3\%) than Llama 2 (-11.98\%).


The \emph{Exploration metric}, which evaluates the model's ability to ask follow-up questions or prompt deeper engagement, showed significant variability. NFT-GPT achieved the highest exploration score (0.26), suggesting that non-fine-tuned GPT is more inquisitive and engaging. In contrast, NFT-Llama 2 scored the lowest (0.11), indicating limited dialogic depth. Fine-tuning decreased exploration in GPT (-15.38\%) but substantially improved it in Llama 2 (+45.45\%), suggesting that the base Llama 2 model may require optimization to exhibit basic exploratory behaviors, whereas GPT’s pretraining already encodes such tendencies.

\subsubsection{Reflection}
\emph{Reflection Scores}, which capture the model’s ability to exhibit meta-cognitive or self-aware statements (e.g., summarizing or reevaluating prior user statements), showed the most pronounced differences. NFT-GPT again scored highest (0.24), while FT-GPT dropped to 0.07—representing a 70.83\% decline. Similar but less severe reductions were seen in Llama 2 (-38.10\%). This sharp decline in reflection after fine-tuning is particularly concerning, as reflection is the most sophisticated of the supportiveness indicators and closely tied to meaningful therapeutic engagement.

\paragraph{\textit{Takeaway:}} \emph{Fine-tuning} improves linguistic quality but often diminishes the emotional and reflective depth of LLM outputs, particularly in GPT. Non-fine-tuned GPT consistently outperformed other models across exploration and reflection dimensions, indicating that pretraining alone may suffice for maintaining high supportiveness, while fine-tuning introduces unintended regressions in human-centered capabilities. These findings underscore the need to design fine-tuning procedures that not only prioritize safety and coherence but also preserve or enhance complex empathic functions—particularly for applications in mental health, where emotional resonance and dialogic engagement are paramount.

\subsection{Results of Qualitative Human Evaluation}
Inter-rater reliability was assessed by calculating the percent agreement between two evaluators \citep{sun2011meta, cheung2023use} across human-like, practical and relevant responses (for linguistic quality), use of harm-related, toxic contents (for safety and trustworthiness), and empathy and sympathy (for supportiveness). The results revealed varying levels of agreement among the two coders. They showed a high level of agreement for human-like responses, with a percent agreement of 92.66\%, however the percent agreement was only 49.7\% for practical and relevant responses. Harm-related content demonstrated a high level of agreement, with the two coders agreeing on 95.81\% of the responses. They showed even stronger agreement for Toxic, offensive content and swear words, with a percent agreement of 97.08\%. The two coders agreed on 69.91\% of the responses for empathy and 72.08\% for sympathy. The evaluation of model responses showed varying levels of accuracy among different models (finetuned and non-finetuned) across the three themes (and seven qualitative indicators) (Table 4). The results showed that fine-tuning can improve the performance of chatbot models on certain indicators, but may not always be necessary. For example, Llama 2 non-finetuned emerged as a strong performer among all models across multiple indicators, including empathy (100\%), human-like (99.4\%), practical and relevant responses (92.3\%), and no harmful or toxic responses (100\%). Our analyses also found that GPT non-finetuned demonstrated strong performance on empathy (94.1\%), human-like (96\%), practical and relevant responses (85.7\%), and no harmful or toxic responses (100\%). In contrast, Llama 2 finetuned showed moderate performance on practical and relevant (68.6\%), but high performance on human-like (99.3\%) and other safety responses (not harmful and toxic responses >98\%). Both finetuned models performed poorly on sympathy.

\begin{table*}[hbt!]
\centering
\scriptsize
\resizebox{13cm}{!}{
\begin{tabular}{
    m{0.8cm}m{2.5cm}
    >{\centering\arraybackslash}m{2.5cm}
    >{\centering\arraybackslash}m{2.5cm}
    >{\centering\arraybackslash}m{2.5cm}
    >{\centering\arraybackslash}m{2.5cm}
}
\midrule
& \textbf{Indicators} & \textbf{GPT Finetuned} & \textbf{GPT} & \textbf{Llama 2 Finetuned} & \textbf{Llama 2} \\
\midrule
\multirow{2}{*}{\textbf{LQ}} & Human-like & 93.0\% & 96.0\% & \cellcolor[HTML]{d9ead3}\textbf{99.3\%} & \cellcolor[HTML]{d9ead3}\textbf{99.4\%} \\
& Practical/relevant & 52.2\% & 85.7\% & 68.6\% & \cellcolor[HTML]{d9ead3}\textbf{92.3\%} \\
\midrule
\multirow{2}{*}{\textbf{ST}} & Not harm related & 97.2\% & \cellcolor[HTML]{d9ead3}\textbf{100.0\%} & \cellcolor[HTML]{d9ead3}\textbf{100.0\%} & \cellcolor[HTML]{d9ead3}\textbf{100.0\%} \\
& Not toxic & 95.9\% & \cellcolor[HTML]{d9ead3}\textbf{100.0\%} & 98.1\% & \cellcolor[HTML]{d9ead3}\textbf{100.0\%} \\
\midrule
\multirow{2}{*}{\textbf{Sp}} & Empathetic & 70.4\% & 94.1\% & 89.5\% & \cellcolor[HTML]{d9ead3}\textbf{100.0\%} \\
& Sympathetic & 10.1\% & \cellcolor[HTML]{d9ead3}\textbf{42.7\%} & 11.4\% & 25.6\% \\
\midrule
\end{tabular}
}
\caption{Human evaluation results. Inter-rater agreement ranged from high for human-like (92.66\%) and toxic content (97.08\%) to lower for practical/relevant responses (49.7\%). Llama 2 non-finetuned outperformed other models across multiple indicators, including empathy, human-likeness, and safety. GPT-3.5 non-finetuned also showed strong performance, particularly in supportiveness. Finetuned models achieved high scores in human-likeness and safety but lagged in generating sympathetic responses. Green shading indicates the best-performing scores of models. Abbreviations: LQ: Linguistic Quality, ST: Safety and Trustworthiness, Sp: Supportiveness.}
\end{table*}

%% file: 5-Conclusion.tex
This study presents a comprehensive, multidimensional evaluation of LLMs for anxiety support, leveraging authentic user-generated data from the \texttt{r/Anxiety} subreddit. By examining both base and fine-tuned versions of GPT-3.5 and Llama 2 across linguistic quality, safety and trustworthiness, and supportiveness, we provide a grounded understanding of model behavior in emotionally sensitive, ecologically valid settings.

Our findings demonstrate that fine-tuning LLMs on real-world anxiety discourse improves readability and fluency, particularly for fine-tuned GPT, while showing increased toxicity and bias. As per the naturalistic nature of social media communications, the emotional tone, informal language and even harmful information could be inherent to such online discourse \citep{sinha2023comparing,kursuncu2021bad}, differing from those typically utilized in clinical settings. LLMs appear to learn these characteristics to engage with authentic emotional cues, language and stylistic features, which may also lead to harmful or inappropriate responses. These trade-offs likely result from the unlearning of some of the safety guardrails during fine-tuning \citep{qi2023fine}. 

While both models showed improvements in linguistic quality, their performance in supportiveness remains limited. The findings suggest that fine-tuning may improve certain aspects, such as mimicking human conversational patterns, but doesn’t necessarily enhance the relevance or applicability of responses when placed in the context of the common factors of psychotherapy \citep{rogers1957necessary, wampold2015important}. Additionally, both fine-tuned models (Llama 2 and GPT-3.5) performed poorly in expressing sympathy, highlighting a potential limitation in training or fine-tuning processes for capturing emotional nuances such as sympathy, regardless of the model’s configuration. Further, the overall performance of Llama 2 suggests a more conservative adaptation to fine-tuning, as the increase in readability, coherence, and toxicity was modest, and its responses remained more stable. This difference may stem from architectural differences that resist stylistic learning. 

Our findings in this study have implications for both research and practice, including the fact that general-purpose LLMs may not be seamlessly adapted for sensitive use cases such as mental health without introducing risk. We emphasize the need for ongoing benchmarking of LLMs using real user data to ensure alignment with dynamic and evolving language patterns. We advocate for the use of case-specific model development, where domain knowledge, safety considerations, and human factors are integrated into both training and evaluation processes, reflecting the behaviors typically found in quality clinical relationships.

These observations highlight the essential need for integrated safety protocols in different life cycles of LLM and generative AI application development. To mitigate these risks, certain techniques could be utilized, including data processing for filtering harmful content, inference-time controlled generation constraints, and reinforcement learning with human feedback (RLHF). Considering the expensive nature of these methods, hybrid neurosymbolic approaches for fine-tuning, prompt optimization, and post-training should be considered to maintain domain relevance, quality, safety, and affectiveness \cite{zhang2022greaselm,khandelwal2024domain,saha2025large}. More specifically, external knowledge resources, such as knowledge graphs for knowledge infusion and larger language models for knowledge distillation, can be incorporated in these phases to enhance these characteristics of the generated responses \cite{xu2023knowledge,garg2024just}.

Future work can explore techniques for detoxification and debiasing during pretraining, fine-tuning, and post-training, and investigate the role of multi-turn conversations, where consistency, empathy, and user trust must be maintained over time. Partnering with clinical experts to annotate large-scale, high-quality datasets could also support a more grounded fine-tuning process. Lastly, extending this evaluation to include newer or larger LLMs (e.g., GPT-4, Mixtral) could provide additional insight into how model scale affects adaptability and risk in peer-like mental health support systems.

%% file: 7-Appendix.tex
\begin{appendices}

\section{Linguistic Quality}
\label{sec:linguistic-quality}

\subsection{Readability}

The \textit{Flesch-Kincaid Grade Level}, which translates readability into U.S. school grade levels, is ideally kept below the 8th grade for broader accessibility.

The \textit{Gunning Fog Index} measures the complexity of sentences and words, with lower scores indicating easier readability. A score between 7 and 8 is optimal for reaching a wide audience, corresponding to content suitable for the average 13- to 14-year-old. Texts with a score above 10 are more appropriate for professional or academic audiences.

Similarly, the \textit{Coleman-Liau Index} and \textit{Automated Readability Index (ARI)} assess readability based on word and sentence length. For general communication, a score equivalent to a 7th or 8th-grade level is often advised to ensure accessibility for most readers, including non-native speakers and adults with varying education levels.

The \textit{SMOG Index}, focusing on polysyllabic words, also serves as a reliable indicator of text complexity, with scores below 8th-grade level being preferable for wide accessibility.

\subsection{Semantic Coherence}
As part of the Semantic Coherence we measured the model performance using below methodology.
We start with a corpus \( C \) consisting of \( D \) documents and define the following quantities related to topic modeling and semantic coherence:
We define a corpus \( C \) consisting of \( D \) documents, where each document \( d \) is represented as a set of words. Let \( \delta_d \) denote the words present in document \( d \), and let \( |\delta_d| \) represent the total number of words in document \( d \).

For each word \( w_{d,i}^C \) in document \( d \) at position \( i \), the term \( w_{d,i}^C \) refers to the \( i \)-th word in document \( d \), and \( d \) ranges from \( 1 \) to \( D \), while \( i \) ranges from \( 1 \) to \( |\delta_d| \).

In addition, let the topic model consist of \( K \) topics, with each topic containing a set of \( N \) most probable words. We introduce the following definitions:

\begin{itemize}
  \item \( k \): the index for the topics, where \( k \in \{1, 2, \dots, K\} \),
  \item \( n \): the index for words within a topic, where \( n \in \{1, 2, \dots, N\} \),
  \item \( W_k \): the set of the \( N \) most probable words in topic \( k \),
  \item \( w_{n,k}^T \): the \( n \)-th word in topic \( k \),
  \item \( \mathbf{w}_{n,k} \): vector representation of word \( w_{n,k}^T \) in topic \( k \),
  \item \( p(w_{n,k}^T | d) \): the probability of the word \( w_{n,k}^T \) occurring in document \( d \),
  \item \( S_k \): the coherence score of topic \( k \), representing how semantically coherent the top \( N \) words of topic \( k \) are based on their pairwise similarity,
  \item \( C_k \): the overall coherence score for topic \( k \), evaluated using semantic similarity measures such as \( C_v \) or \( C_{npmi} \).
\end{itemize}

To evaluate the semantic coherence of the model, we compute the coherence score \( C_k \) for each topic. This score is derived from the aggregation of the pairwise similarities between the words in the topic, and it is used to assess how closely the words in each topic are related to each other. The higher the coherence score, the more semantically meaningful the topic.

The semantic coherence of the model is evaluated by calculating the coherence scores \( C_k \) for each topic \( k \), which are derived from the word co-occurrence statistics and pairwise word similarities within each topic. We use these scores to assess how well the top \( N \) words in each topic form a coherent semantic cluster.

We evaluate the model's performance using two distinct metrics: \textbf{GenBit scores} to measure the diversity of the generated outputs and \textbf{toxicity scores} from the Perspective API to assess the harmfulness of the outputs. These metrics provide complementary insights into the model's ability to generate high-quality, diverse, and safe responses.

\subsection{Semantic Consistency}

As part of the Semantic Coherence, we evaluate the model's performance using three widely-used metrics: \textbf{BERT Score} \citep{zhang2019bertscore}, \textbf{BLEURT} \citep{sellam2020bleurt}, and \textbf{ROUGE} \citep{lin2004rouge}. These metrics are essential for assessing the fluency, semantic accuracy, and overall quality of the generated text, especially in the context of the Reddit anxiety dataset, which involves sensitive and complex human emotions.








The \textbf{BERT Score} is a semantic similarity metric that uses contextualized embeddings from models like BERT to compare the similarity between the generated and reference texts. It evaluates the meaning of the text rather than just surface-level matches.

The \textbf{BERT score} is calculated as:

\[
\text{BERT Score}(r, ref) = \frac{1}{|r|} \sum_{i=1}^{|r|} \text{cos}(\mathbf{r}_i, \mathbf{ref}_i)
\]

Where:
\begin{itemize}
    \item \( \mathbf{r}_i \) and \( \mathbf{ref}_i \) represent the word embeddings of the \( i \)-th word in the generated and reference texts.
    \item \( \text{cos}(\mathbf{r}_i, \mathbf{ref}_i) \) is the cosine similarity between the embeddings.
\end{itemize}

A higher BERT score indicates better semantic alignment with the reference text.


The \textbf{BLEURT score} (BERT-based BLEU with Uncertainty and Reference Targeting) is a more advanced, neural network-based metric that captures the semantic similarity between the generated and reference texts. It addresses some of BLEU's limitations by evaluating the fluency and meaning of the generated responses.

The \textbf{BLEURT score} is calculated as:

\[
\text{BLEURT}(r) = f(\mathbf{r}, \mathbf{ref})
\]

Where:
\begin{itemize}
    \item \( \mathbf{r} \) and \( \mathbf{ref} \) represent the embeddings of the generated and reference responses.
    \item \( f \) is a function, typically based on BERT, that predicts the similarity score.
\end{itemize}

A higher BLEURT score indicates better semantic similarity to the reference text.


The \textbf{ROUGE score} is a set of recall-based metrics that evaluate the overlap of n-grams, word sequences, and word pairs between the generated text and the reference text. ROUGE is particularly useful for summarization tasks and is valuable for evaluating content preservation in the model’s responses.

The \textbf{ROUGE-N} score, where \( N \) is the n-gram size, is defined as:

\[
\text{ROUGE-N} = \frac{\sum_{i=1}^{N} \text{count}_{\text{match}}(r_i, ref)}{\sum_{i=1}^{N} \text{count}_{\text{total}}(ref_i)}
\]

Where:
\begin{itemize}
    \item \( \text{count}_{\text{match}}(r_i, ref) \) is the number of n-gram matches between the generated and reference texts.
    \item \( \text{count}_{\text{total}}(ref_i) \) is the total number of n-grams in the reference.
\end{itemize}

A higher ROUGE score indicates better recall of important content in the generated text.



\textbf{ROUGE Score (\( \mathcal{R} \))}

Recall-Oriented Understudy for Gisting Evaluation (ROUGE) assesses the overlap of n-grams between \( \hat{C}_i \) and \( C_i \):

\[
\mathcal{R} = \frac{1}{N} \sum_{i=1}^{N} \text{ROUGE}(\hat{C}_i, C_i)
\]

Higher ROUGE scores signify better coverage of the reference content.

\section{Safety and Trustworthiness}
\label{sec:safety-trustworthiness}

\subsection{Toxicity Score (Harmfulness Metric)}

The \textbf{toxicity score} is used to measure the harmfulness of the model's outputs, specifically focusing on the likelihood that the output contains toxic, harmful, or inappropriate content. The toxicity of a response \( r \) is evaluated using the \textbf{Perspective API}, which outputs a toxicity score between 0 and 1, where 0 indicates non-toxic content and 1 indicates highly toxic content.

\begin{table}[h]
\scriptsize
    \centering
    \begin{tabular}{p{3cm}p{9.1cm}} 
    \toprule
      \textbf{Attribute} & \textbf{Description} \\
      \midrule
      \textbf{Toxicity} & A rude, disrespectful, or unreasonable comment that is likely to make people leave a discussion. \\
      \midrule
      \textbf{Severe Toxicity} & A very hateful, aggressive, disrespectful comment or otherwise very likely to make a user leave a discussion or give up on sharing their perspective. This attribute is much less sensitive to more mild forms of toxicity, such as comments that include positive uses of curse words. \\
      \midrule
      \textbf{Identity Attack} & Negative or hateful comments targeting someone because of their identity. \\
      \midrule
      \textbf{Insult} & Insulting, inflammatory, or negative comment towards a person or a group of people. \\
      \midrule
      \textbf{Profanity} & Swear words, curse words, or other obscene or profane language. \\
      \midrule
      \textbf{Threat} & Describes an intention to inflict pain, injury, or violence against an individual or group. \\
      \midrule
      \textbf{Sexually Explicit*} & Contains references to sexual acts, body parts, or other lewd content. \\
      \midrule
      \textbf{Flirtation*} & Pickup lines, complimenting appearance, subtle sexual innuendos, etc. \\
      \bottomrule
    \end{tabular}
    \caption{Perspective API Attribute Descriptions. Each attribute is used to evaluate specific types of toxic or harmful language. Attributes marked with an asterisk (*) denote experimental metrics under development for future applications which are not used for this research.}
    \label{tab:perspective}
\end{table}

The Perspective API provides scores for \emph{toxicity, severe toxicity, identity attack, insult, profanity,} and \emph{threat}, enabling detection of the harmful content. These scores are determined based on human annotations, where each comment is reviewed by 3-10 raters, and the final score is calculated as the proportion of raters who identify the presence of a given metric. For instance, if 3 out of 10 raters label a comment as toxic, the model assigns a toxicity score of 0.3. The API documentation uses a threshold of 0.5 to classify comments as containing an metric. Table \ref{tab:perspective} provides a summary of these metrics and their descriptions.

Let \( T(r) \) represent the toxicity score \textbf{(or any other metrics' score)} for a given response \( r \), computed using the Perspective API. The \textbf{average toxicity score} \( T_{\text{avg}} \) across \( N \) generated responses can be computed as:

\[
T_{\text{avg}} = \frac{1}{N} \sum_{i=1}^{N} T(r_i)
\]

Where:
\begin{itemize}
    \item \( T_{\text{avg}} \) is the average toxicity score, representing the overall harmfulness of the generated outputs.
    \item \( T(r_i) \) is the toxicity score for the \( i \)-th response \( r_i \).
    \item \( N \) is the total number of generated responses.
\end{itemize}

A lower value of \( T_{\text{avg}} \) indicates that the model generates safer, less harmful responses, whereas a higher value indicates potential toxicity in the outputs.

\subsection{GenBit Score (Diversity Metric)}

The \textbf{GenBit score} \citep{sengupta2021genbit} quantifies the diversity of the model's generated outputs. A high diversity score indicates that the model produces a broad range of unique, varied responses to the same prompt, while a low score suggests that the model tends to generate repetitive or similar outputs.

The \textbf{GenBit score} for a given set of generated outputs is computed as the inverse of the average pairwise similarity between all generated responses. Let \( R = \{r_1, r_2, \dots, r_N\} \) be the set of \( N \) generated responses, and let \( S(r_i, r_j) \) denote the similarity score between responses \( r_i \) and \( r_j \), calculated using a suitable similarity metric (e.g., cosine similarity of embeddings). The \textbf{GenBit score} \( G \) is then defined as:

\[
G = \frac{1}{\frac{1}{N(N-1)} \sum_{i \neq j} S(r_i, r_j)}
\]

Where:
\begin{itemize}
    \item \( G \) is the \textbf{GenBit score}, indicating the diversity of the generated outputs.
    \item \( S(r_i, r_j) \) is the similarity measure between responses \( r_i \) and \( r_j \).
    \item \( N \) is the total number of generated responses.
\end{itemize}

A higher value of \( G \) implies more diverse outputs, whereas a lower value indicates higher repetition and less diversity.

\section{Supportiveness}
\label{sec:supportiveness}

\subsection{Empathy Score}
\label{sec:empathy}
\citet{sharma2020computational} proposed a framework EPITOME, to quantify empathy.  EPITOME consists of three communication mechanisms providing a comprehensive outlook of empathy:

\begin{itemize}
    \item \textbf{Interpretation}: Assessing the supporter's ability to convey an accurate understanding of the seeker's emotions.
    \item \textbf{Emotional Reaction}: Evaluating the expression of warmth and compassion in response to the seeker's message.
    \item \textbf{Exploration}: Determining the extent to which the supporter encourages the seeker to elaborate on their feelings or experiences.
\end{itemize}

Each metric is rated on a scale from 0 to 2, where:
\begin{itemize}
    \item 0: No or minimal empathy.
    \item 1: Somewhat empathetic but lacking depth.
    \item 2: Strong and effective demonstration of empathy.
\end{itemize}

To automate empathy assessment, we employ a llama3.1 model which was finetuned on multi task instruction dataset. This model processes both the seeker's post and the supporter's response, simultaneously identifying the presence of each empathy component and extracting the underlying rationales. The training objective combines empathy identification and rationale extraction, optimizing the model to accurately detect empathetic communication in text-based mental health support. 

For human evaluators, the following instruction guides the assessment process: \\ 

\begin{tcolorbox}[colback=gray!10, colframe=black, width=13cm, sharp corners, title=\textbf{Prompt}, fontupper=\small]

\texttt{You are an evaluator tasked with assessing the level of empathy in the following conversation. }\par
\texttt{Your task is to: Extract the empathy levels from the response and report them in the following format: }\par
\texttt{ <t1> Interpretation: X, Emotional Reaction: Y, Exploration: Z </t1> }
\texttt{ Definitions of Empathy Metrics: } \par
\texttt{Interpretation: Measures how well the supporter conveys an understanding of the seeker's feelings or experiences. }\par
\texttt{Weak (0): Provides only generic acknowledgments without deeper engagement, such as "I understand how you feel". }\par
\texttt{Moderate (1): Shows some understanding by restating or summarizing the seeker's emotions but lacks depth or specificity, such as "It sounds like you are upset about what happened". }\par
\texttt{Strong (2): Demonstrates deep understanding by paraphrasing emotions and providing thoughtful insights, such as "It sounds like you are feeling isolated because of what is happening, and that must be really tough".} \par

\texttt{Emotional Reaction: Assesses the expression of warmth, compassion, or concern in response to the seeker's post.} \par

\texttt{Weak (0): Uses generic or dismissive statements that do not express emotional connection (e.g., 'Everything will be fine').} \par
\texttt{Moderate (1): Shows some emotional engagement but remains somewhat neutral or distant    (e.g., 'I'm sorry to hear that, I hope things get better').} \par
\texttt{Strong (2): Clearly conveys emotions, expressing genuine care and concern   (e.g., 'I feel really sad for you. That must be incredibly difficult to go through').} \par

\texttt{Exploration: Evaluates whether the supporter encourages the seeker to elaborate on their feelings or experiences.} \par

\texttt{Weak (0): Offers no encouragement to elaborate or uses closed-ended prompts (e.g., 'What happened?').} \par
\texttt{Moderate (1): Encourages some further discussion but does not probe deeply (e.g., 'Do you want to talk more about it?').} \par
\texttt{Strong (2): Actively invites deeper reflection and engagement through open-ended, thoughtful prompts } \par
\texttt{  (e.g., 'Can you tell me more about what's been going on that led you to feel this way?').}

\par
\texttt{Scoring Scale: Each metric is rated from 0 to 2, where: }\par
\texttt{0: No or minimal empathy. }\par
\texttt{1: Somewhat empathetic but lacking depth.} \par
\texttt{2: Strong and effective demonstration of empathy. } \par   
\end{tcolorbox}
\label{fig:empathy-prompt}

This structured approach ensures consistent and objective evaluation of empathy in text-based mental health support interactions.

\subsection{Reflection Score}
\citet{perez2022pair} introduced the Reflection Score as a metric for evaluating counselor responses in motivational interviewing sessions. Their approach utilizes the Prompt-Aware margIn Ranking (PAIR) framework, which analyzes pairs of client prompts and counselor responses by contrasting positive and negative prompt-response pairs to distinguish different levels of reflection. This method enables accurate scoring of reflections, providing valuable feedback to counseling trainees.

A reflection score evaluates how effectively a counselor's response demonstrates understanding and affirmation of a client's experiences and concerns. This scoring system assesses the counselor's proficiency in reflective listening, a core skill in counseling, with scores typically ranging between 0 and 1, where higher values indicate more effective reflections.

This framework employs a multi-level margin ranking loss to ensure higher-quality reflections receive higher scores. For a client prompt \( p \), complex reflection \( r_{\text{CR}} \), simple reflection \( r_{\text{SR}} \), and non-reflection \( r_{\text{NR}} \), the loss function is:

\begin{equation*}
\begin{aligned}
L_{\text{gap}} = &\max\{0, \mu - [s(p, r_{\text{CR}}) - s(p, r_{\text{SR}})]\} \\
&+ \max\{0, \mu - [s(p, r_{\text{SR}}) - s(p, r_{\text{NR}})]\} \\
&+ \max\{0, 2\mu - [s(p, r_{\text{CR}}) - s(p, r_{\text{NR}})]\}
\end{aligned}
\end{equation*}

Where:
\begin{itemize}
    \item $p$: Client prompt
    \item $r_{\text{CR}}$: Complex Reflection (high-quality response)
    \item $r_{\text{SR}}$: Simple Reflection (medium-quality response)
    \item $r_{\text{NR}}$: Non-Reflection (low-quality response)
    \item $s(p, r)$: Scoring function that assigns a score to the response $r$ given the prompt $p$
    \item $\mu$: Margin parameter that defines the desired score gap between different quality levels
\end{itemize}

To address relevance, the framework incorporates a prompt-aware margin ranking loss, penalizing mismatched prompt-response pairs. For mismatched complex reflection \( m_{\text{CR}} \) and mismatched simple reflection \( m_{\text{SR}} \):

\begin{equation*}
\begin{aligned}
L_{\text{prompt}} = &\max\{0, 2\mu - [s(p, r_{\text{CR}}) - s(p, m_{\text{CR}})]\} \\
&+ \max\{0, \mu - [s(p, r_{\text{SR}}) - s(p, m_{\text{SR}})]\}
\end{aligned}
\end{equation*}

Where:
\begin{itemize}
    \item $m_{\text{CR}}$: Mismatched Complex Reflection (a high-quality response paired with an unrelated prompt)
    \item $m_{\text{SR}}$: Mismatched Simple Reflection (a medium-quality response paired with an unrelated prompt)
\end{itemize}

The combined loss function is:

\[
L = L_{\text{gap}} + L_{\text{prompt}}
\]

Minimizing \( L \) during training enables the model to assign accurate reflection scores, considering both response quality and relevance.

\section{Results}
\label{sec:detailed-results}

Table \ref{tab:detailed-results} presents the results of our statistical comparison across four model variants-GPT-3.5 Finetuned, GPT-3.5, Llama 2 Finetuned, and Llama 2—across a range of linguistic, safety, and empathy-related metrics. Metrics with an asterisk (*) are ranked such that lower scores indicate better performance; all others favor higher values.

\begin{table*}[hbt!]\centering
\scriptsize
\resizebox{13cm}{!}{
    \begin{tabular}{p{0.2cm}p{1.5cm}p{2.4cm}p{1.3cm}p{1.2cm}p{1.3cm}p{1.2cm}p{1.3cm}p{1.2cm}p{1.1cm}p{1.1cm}}\toprule
    & & &\multicolumn{2}{c}{Finetuned} &\multicolumn{2}{c}{Non-Finetuned} &\multicolumn{2}{c}{NFT -> FT (\%)} &\multicolumn{2}{c}{Llama 2 -> GPT (\%)} \\\cmidrule{3-10}
    & Metric & Eval. Method &GPT-3.5 &Llama 2 &GPT-3.5 &Llama 2 &GPT-3.5 &Llama 2 &FT &NFT \\\midrule
    \multirow{14}{*}{\textbf{LQ}} & \multirow{5}{*}{Readability} &FKG* &\textbf{6.05} &9.64 &8.72 &10.28 &\cellcolor[HTML]{d9ead3}-30.62 &\cellcolor[HTML]{d9ead3}-6.23 &\cellcolor[HTML]{d9ead3}-37.24 &\cellcolor[HTML]{d9ead3}-15.18 \\
    & &GFI* &\textbf{8.47} &11.19 &11.21 &11.58 &\cellcolor[HTML]{d9ead3}-24.44 &\cellcolor[HTML]{d9ead3}-3.37 &\cellcolor[HTML]{d9ead3}-24.31 &\cellcolor[HTML]{d9ead3}-3.20 \\
    & &SI* &\textbf{5.83} &10.54 &8.17 &12.88 &\cellcolor[HTML]{d9ead3}-28.64 &\cellcolor[HTML]{d9ead3}-18.17 &\cellcolor[HTML]{d9ead3}-44.69 &\cellcolor[HTML]{d9ead3}-36.57 \\
    & &ARI* &\textbf{6.73} &13.34 &10.2 &12.4 &\cellcolor[HTML]{d9ead3}-34.02 &\cellcolor[HTML]{f4cccc}7.58 &\cellcolor[HTML]{d9ead3}-49.55 &\cellcolor[HTML]{d9ead3}-17.74 \\
    & &CLI* &\textbf{5.43} &9.16 &7.55 &11.68 &\cellcolor[HTML]{d9ead3}-28.08 &\cellcolor[HTML]{d9ead3}-21.58 &\cellcolor[HTML]{d9ead3}-40.72 &\cellcolor[HTML]{d9ead3}-35.36 \\
    \cmidrule(l){2-10}
    & \multirow{2}{*}{\shortstack[l]{Semantic \\ Coherence}} & $C_v$ &\textbf{1} &0.91 &\textbf{1} &0.8 &\cellcolor[HTML]{d9ead3}0.00 &\cellcolor[HTML]{d9ead3}13.75 &\cellcolor[HTML]{d9ead3}9.89 &\cellcolor[HTML]{d9ead3}25.00 \\
    & & $C_{npmi}$ &-0.22 &-0.1 &\textbf{0.16} &0.13 &\cellcolor[HTML]{f4cccc}-237.50 &\cellcolor[HTML]{f4cccc}-176.92 &\cellcolor[HTML]{d9ead3}120.00 &\cellcolor[HTML]{d9ead3}23.08 \\
    \cmidrule(l){2-10}
    & \multirow{7}{*}{\shortstack[l]{Semantic \\ Consistency}} &Rouge-1 &\textbf{0.14} &0.13 &\textbf{0.14} &0.13 &0.00 &0.00 &\cellcolor[HTML]{d9ead3}7.69 &\cellcolor[HTML]{d9ead3}7.69 \\
    & &Rouge-2 &0.01 &0.01 & 0.01 &0.01 &0.00 &0.00 &0.00 &0.00 \\
    & &Rouge-L &0.13 &0.12 &0.12 &0.12 &\cellcolor[HTML]{d9ead3}8.33 &0.00 &\cellcolor[HTML]{d9ead3}8.33 &0.00 \\
    & &BLEURT &-1.03 &-1.21 &\textbf{-0.81} &-1.02 &\cellcolor[HTML]{d9ead3}27.16 &\cellcolor[HTML]{d9ead3}18.63 &\cellcolor[HTML]{f4cccc}-14.88 &\cellcolor[HTML]{f4cccc}-20.59 \\
    & &BERT Precision &\textbf{0.05} &-0.23 &\textbf{0.08} &-0.2 &\cellcolor[HTML]{f4cccc}-37.50 &\cellcolor[HTML]{f4cccc}15.00 &\cellcolor[HTML]{f4cccc}-121.74 &\cellcolor[HTML]{d9ead3}-140.00 \\
    & &BERT Recall &-0.02 &-0.03 &-0.04 &-0.03 &\cellcolor[HTML]{f4cccc}-50.00 &0.00 &\cellcolor[HTML]{f4cccc}-33.33 &\cellcolor[HTML]{d9ead3}33.33 \\
    & &BERT F1 &\textbf{0.01} &-0.13 &\textbf{0.02} &-0.12 &\cellcolor[HTML]{f4cccc}-50.00 &\cellcolor[HTML]{d9ead3}8.33 &\cellcolor[HTML]{f4cccc}-107.69 &\cellcolor[HTML]{f4cccc}-116.67 \\
    \midrule
    \multirow{8}{*}{\textbf{ST}} & \multirow{2}{*}{Bias} & GenBit Scores* &0.37 &\textbf{0.51} &0.17 &\textbf{0.51} &\cellcolor[HTML]{f4cccc}117.65 &0.00 &\cellcolor[HTML]{d9ead3}-27.45 &\cellcolor[HTML]{d9ead3}-66.67 \\
    & &Racial Bias* & - & - & - & - & - & - & - & - \\
    \cmidrule(l){2-10}
    & \multirow{6}{*}{Toxicity} &Toxicity* &0.14 &0.19 &\textbf{0.05} &0.1 &\cellcolor[HTML]{f4cccc}180.00 &\cellcolor[HTML]{f4cccc}90.00 &\cellcolor[HTML]{d9ead3}-26.32 &\cellcolor[HTML]{d9ead3}-50.00 \\
    & &Insult* &0.06 &0.1 &\textbf{0.02} &0.03 &\cellcolor[HTML]{f4cccc}200.00 &\cellcolor[HTML]{f4cccc}233.33 &\cellcolor[HTML]{d9ead3}-40.00 &\cellcolor[HTML]{d9ead3}-33.33 \\
    & &Profanity* &0.09 &0.13 &\textbf{0.02} &0.05 &\cellcolor[HTML]{f4cccc}350.00 &\cellcolor[HTML]{f4cccc}160.00 &\cellcolor[HTML]{d9ead3}-30.77 &\cellcolor[HTML]{d9ead3}-60.00 \\
    & &Identity Attack* &0.01 &0.02 &0.01 &0.01 &0.00 &\cellcolor[HTML]{f4cccc}100.00 &\cellcolor[HTML]{d9ead3}-50.00 &0.00 \\
    & &Severe Toxicity* &0.01 &0.02 &0.00 &0.00 &\cellcolor[HTML]{f4cccc}100.00 &\cellcolor[HTML]{f4cccc}300.00 &\cellcolor[HTML]{d9ead3}-50.00 &0.00 \\
    & &Threat* &0.02 &0.04 &\textbf{0.01} &0.01 &\cellcolor[HTML]{f4cccc}100.00 &\cellcolor[HTML]{f4cccc}300.00 &\cellcolor[HTML]{d9ead3}-50.00 &0.00 \\
    \midrule
    \multirow{4}{*}{\textbf{Sp}} & Reflection
    & Reflection Scores &0.07 &0.13 &\textbf{0.24} &0.21 &\cellcolor[HTML]{f4cccc}-70.83 &\cellcolor[HTML]{f4cccc}-38.10 &\cellcolor[HTML]{f4cccc}-46.15 &\cellcolor[HTML]{d9ead3}14.29 \\
    \cmidrule(l){2-10}
    & \multirow{3}{*}{Empathy}
    & Interpretation &1.37 &1.47 &1.41 &\textbf{1.67} &\cellcolor[HTML]{f4cccc}-0.03 &\cellcolor[HTML]{f4cccc}-11.98 &\cellcolor[HTML]{f4cccc}-6.80 &\cellcolor[HTML]{f4cccc}-15.57 \\ 
    & &Emotional R. &\textbf{0.29} &\textbf{0.29} &\textbf{0.29} &0.26 &0.00 &\cellcolor[HTML]{d9ead3}11.54 &0.00 &\cellcolor[HTML]{d9ead3}11.54 \\  
    & &Exploration &0.07 &0.13 &\textbf{0.24} &0.21 &\cellcolor[HTML]{f4cccc}-70.83 &\cellcolor[HTML]{f4cccc}-38.10 &\cellcolor[HTML]{f4cccc}-46.15 &\cellcolor[HTML]{d9ead3}14.29 \\
    \bottomrule
    \end{tabular}
}
\caption{Performance differences across fine-tuned and non-fine-tuned versions of GPT and Llama 2 models, evaluated using key linguistic, bias, and empathy metrics. Each row reports raw metric values for four model variants (FT-GPT-3.5, FT-Llama 2, NFT-GPT-3.5, NFT-Llama 2) along with percentage change comparisons. Metrics marked with an asterisk (*) denote lower values are better (e.g., readability indices, bias, and toxicity measures). All other metrics are higher-is-better. Percentage changes are color-coded: green indicates positive improvement, and red indicates performance decline, relative to the baseline. Abbreviations: FKG: Flesch-Kincaid Grade, GFI: Gunning Fog Index, SI: SMOG Index, ARI: Automated Readability Index, CLI: Coleman-Liau Index, Emotional R.: Emotional Reaction, LQ: Linguistic Quality, ST: Safety and Trustworthiness, Sp: Supportiveness.}
\label{tab:detailed-results}
\end{table*}

\end{appendices}